\documentclass[aps,prd,onecolumn,superscriptaddress,nofootinbib,showpacs]{revtex4}

\usepackage{graphicx}
\usepackage{slashed}
\usepackage{amsmath,bbm,latexsym,amssymb}
\usepackage{bbold}
\usepackage{epsfig}
\usepackage{epstopdf}
\usepackage{color}

\newcommand{\be}{\begin{equation}}
\newcommand{\ee}{\end{equation}}
\newcommand{\ba}{\begin{eqnarray}}
\newcommand{\ea}{\end{eqnarray}}
\newcommand{\bs}{\begin{subequations}}
\newcommand{\es}{\end{subequations}}

\newcommand{\grts}{\raise.3ex\hbox{$>$\kern-.75em\lower1ex\hbox{$\sim$}}}
\newcommand{\lets}{\raise.3ex\hbox{$<$\kern-.75em\lower1ex\hbox{$\sim$}}}

\begin{document}
%\begin{flushright}
%CFTP-10/01\\[-1mm]
%%arXiv:????.???? [hep-ph]\\[-1mm]
%\end{flushright}
\vspace*{1cm}

\title{Tree-level metastability bounds for the most general two Higgs doublet model}

\author{I. P. Ivanov}\thanks{E-mail: igor.ivanov@tecnico.ulisboa.pt}
\affiliation{CFTP, Departamento de F\'{\i}sica,
Instituto Superior T\'{e}cnico, Universidade de Lisboa,
Avenida Rovisco Pais 1, 1049 Lisboa, Portugal}
\author{Jo\~{a}o P.\ Silva}\thanks{E-mail: jpsilva@cftp.ist.utl.pt}
\affiliation{CFTP, Departamento de F\'{\i}sica,
Instituto Superior T\'{e}cnico, Universidade de Lisboa,
Avenida Rovisco Pais 1, 1049 Lisboa, Portugal}

\date{\today}

\begin{abstract}
Within two Higgs doublet models, it is possible that the current
vacuum is not the global minimum, in which case it could possibly
decay at a later stage.
We discuss the tree-level conditions which must be obeyed by the
most general scalar potential in order to preclude that possibility.
We propose a new procedure which is not only more general but also
easier to implement than the previously published one,
including CP conserving as well as CP violating scalar sectors.
We illustrate these conditions within the context of the
$Z_2$ model, softly broken by a complex, CP violating parameter.
\end{abstract}


\maketitle

\section{Introduction}

The discovery of a fundamental scalar particle \cite{ATLASHiggs,CMSHiggs}
prompts the search for other fundamental scalars.
The simplest theoretical possibility consists in adding
a second doublet to the Standard Model (SM),
forming a two Higgs doublet model (2HDM) --
for reviews see, for example, Refs.~\cite{hhg,ourreview}.

There are several theoretical constraints that a viable 2HDM scalar sector
must obey:
it must be bounded from below \cite{Deshpande:1977rw},
and it must conform to perturbative unitarity
\cite{Kanemura:1993hm, Akeroyd:2000wc, Ginzburg:2003fe, Ginzburg:2005dt}.
In addition,
the scalar potential of the 2HDM may have simultaneously
two neutral minima,
both CP conserving or both CP-violating
\cite{NagelPhD, Maniatis:2006fs, Ivanov:2006yq, Ivanov:2007de, Ivanov:2008cxa}.
In those cases,
if one were in the local metastable minimum,
there would be the possibility of decaying later
into the global minimum: the true vacuum.
We dub this possibility the \textit{panic vacuum}.

When building a model with a scalar sector allowing for several distinct minima,
one is free to take a metastable minimum as physically acceptable, provided it is sufficiently long-lived,
or to disregard it altogether. But one definitely cannot neglect this issue.
Typically,
this would be investigated by reconstructing the potential from the
input parameters (vacuum expectation values, masses, etc\dots),
finding the second vacuum numerically and comparing the potential depths
in the two vacua.
This is time consuming, especially if one performs extensive scans over parameter space,
and it could give a false positive.
Luckily,
within the 2HDM there is an efficient way to study these issues
without trying to solve the minimization problem.
Following the works \cite{Ivanov:2006yq, Ivanov:2007de, Ivanov:2008cxa},
a discriminant for panic vacua has been studied in the context of
the 2HDM with a softly broken $U(1)$ symmetry \cite{Barroso:2012mj},
which we denote by ``$U(1)$-2HDM'',
and in the context of a $Z_2$-symmetric
2HDM, softly broken by a real parameter \cite{Barroso:2013awa},
which we denote by ``real 2HDM''.
The discriminant in this case is a compact combination
of the parameters of the potential and the vacuum expectation values
(vevs) calculated at a single minimum, whose positive sign guarantees tree-level stability
and negative sign signals the presence of a deeper minimum.

A technique applicable in the case of a scalar potential
without CP violation was also presented in Ref.~\cite{Barroso:2013awa}.
The technique had five steps:
i) perform a numerical search for the eigenvalues of
a certain $4 \times 4$ matrix ($\Lambda_E$);
ii) perform a numerical calculation of its eigenvectors;
iii) combine the eigenvectors into a rotation matrix,
used to transform the parameters of the potential;
iv) determine whether a certain rotated quantity $\hat{M}_0$ were positive or negative;
v) apply the discriminant, distinguishing between the metastable vacuum
and the global minimum (at tree level).

Those studies are not applicable,
for example,
to the 2HDM with a complex soft breaking of the
$Z_2$ symmetry,
usually known as the C2HDM \cite{Ginzburg:2002wt,Khater:2003wq,ElKaffas:2007rq,
El Kaffas:2006nt,WahabElKaffas:2007xd,Osland:2008aw,Grzadkowski:2009iz,Arhrib:2010ju}.
There has been renewed interest in this model,
in part because it is the simplest model allowing for the 125 GeV Higgs
to have a mixture of scalar and pseudoscalar
components \cite{Barroso:2012wz, Shu:2013uua,
Inoue:2014nva, Fontes:2014xva, Fontes:2015mea, Chen:2015gaa}.
Remarkably,
all the data available is still consistent with the
possibility that its couplings to the up quarks are mainly scalar
while its couplings to the down quarks are mainly pseudoscalar
\cite{Fontes:2015mea}.
So far,
no simple criteria to avoid metastable vacua has
been devised for potentials with CP violation,
either spontaneous or explicit.

In this article,
we solve that problem,
thus completing the study of the 2HDM.
We introduce a new discriminant which,
using very simple criteria,
distinguishes between a metastable vacuum and the global minimum.
Besides devising a discriminant applicable to the most general 2HDM,
we also improve upon previous analysis of the $U(1)$-2HDM and the
real 2HDM by presenting the discriminant in terms of physical
masses, mixing angles, and couplings
rather than the parameters of the potential.
In the most general 2HDM, our procedure improves over
the one suggested in \cite{Barroso:2013awa} for the CP conserving
2HDM in two ways.
First, we build it only from eigenvalues of $\Lambda_E$ and do not ask
for the calculation of the eigenvectors
nor for explicit basis transformation.
Second, we prove that our method is safe even in
pathological cases of potentials unbounded from below
with matrix $\Lambda_E$ having complex eigenvalues.

The paper is organized as follows.
In section \ref{sec:notation} we present our notation,
used in section \ref{sec:D} to develop the discriminant
$D$ and the method which disentangles the metastable vacuum from
the true vacuum.
In section \ref{sec:examples},
we discuss in detail the $U(1)$-2HDM,
the real 2HDM,
and the C2HDM,
presenting the discriminant in terms of physical
parameters, such as masses and mixing angles,
instead of parameters of the potential.
In these cases,
there is a simple condition forcing the potential to
be bounded from below,
and the method to avoid panic vacua involves
a single application of a discriminant $\tilde{D}$.
We draw our conclusions in section \ref{sec:concl}
and relegate some detailed proofs to the appendices.

\section{\label{sec:notation}Notation}

The Higgs potential of the most general 2HDM is usually
written as \cite{ourreview}
\ba
V_H
&=&
m_{11}^2 |\phi_1|^2
+ m_{22}^2 |\phi_2|^2
- \left[m_{12}^2\, \phi_1^\dagger \phi_2 +
(m_{12}^2)^\ast\, \phi_2^\dagger \phi_1  \right]
\nonumber\\
&&
+ \frac{\lambda_1}{2} |\phi_1|^4
+ \frac{\lambda_2}{2} |\phi_2|^4
+ \lambda_3 |\phi_1|^2 |\phi_2|^2
+ \lambda_4\, (\phi_1^\dagger \phi_2)\, (\phi_2^\dagger \phi_1)
\nonumber\\
&&
+ \frac{\lambda_5}{2} (\phi_1^\dagger \phi_2)^2
+ \frac{\lambda_5^\ast}{2} (\phi_2^\dagger \phi_1)^2
+ |\phi_1|^2 \left[ \lambda_6 \, \phi_1^\dagger \phi_2
+ \lambda_6^\ast\, \phi_2^\dagger \phi_1 \right]
\nonumber\\
&&
+ |\phi_2|^2 \left[\lambda_7\, \phi_1^\dagger \phi_2
+ \lambda_7^\ast\, \phi_2^\dagger \phi_1 \right],
\label{VH}
\ea
where hermiticity forces all couplings to be real, except $m_{12}^2$ ,
$\lambda_5$, $\lambda_6$, and $\lambda_7$.
If all couplings are real,
then there is explicit CP conservation.

For the study of minima,
an alternative formulation in terms of quantities bilinear in the fields
is useful \cite{NagelPhD, Maniatis:2006fs,
Ivanov:2006yq, Ivanov:2007de, Ivanov:2008cxa,Nishi:2006tg,Nishi:2007nh,Nishi:2007dv}:
\be
V_H =
- M_\mu \mathbb{r}^\mu + \frac{1}{2} \Lambda_{\mu \nu} \mathbb{r}^\mu \mathbb{r}^\nu,
\label{VH_bilinears}
\ee
where\footnote{We adhere here to the original notation of
\cite{Ivanov:2006yq, Ivanov:2007de, Ivanov:2008cxa} inspired by
relativity,
where it is the contravariant vectors which define the spatial components
with the positive sign:
$V^\mu = \left( V_0, V_1, V_2, V_3 \right)$.
This is the opposite of the notation used for
contravariant vectors in Ref.\cite{Barroso:2013awa}.}
\be
\mathbb{r}^\mu
=
\left( \mathbb{r}_0, \mathbb{r}_1, \mathbb{r}_2, \mathbb{r}_3 \right)
=
\left(
|\phi_1|^2 + |\phi_2|^2,\,
2 \textrm{Re} \left( \phi_1^\dagger \phi_2\right),\,
2 \textrm{Im} \left( \phi_1^\dagger \phi_2\right),\,
|\phi_1|^2 - |\phi_2|^2
\right),
\ee
and
\be
M_\mu
=
\left( -{m_{11}^2 + m_{22}^2 \over 2},\,
\textrm{Re}\left( m_{12}^2\right),\,
-\textrm{Im}\left( m_{12}^2\right),\,
-{m_{11}^2 - m_{22}^2 \over 2}
\right),
\label{M}
\ee
while $\Lambda_{\mu \nu}$ is a symmetric matrix.
For our purposes,
it is more useful to consider the mixed symmetry tensor
$\Lambda^\mu_{.\, \nu} = g^{\mu \alpha} \Lambda_{\alpha \nu}$,
viewed as a matrix in Euclidean space:
\be
\Lambda_E = \Lambda^\mu_{.\, \nu}
=
\frac{1}{2}
\left(
\begin{array}{cccc}
	\ \tfrac{1}{2}(\lambda_1 + \lambda_2) + \lambda_3\  &
	\ \textrm{Re}\left( \lambda_6 + \lambda_7 \right)\  &
	\  -\textrm{Im}\left( \lambda_6 + \lambda_7 \right)\  &
	\ \tfrac{1}{2}(\lambda_1 - \lambda_2)\ \\*[1mm]
	\ -\textrm{Re}\left( \lambda_6 + \lambda_7 \right)\  &
	\ - \lambda_4 - \textrm{Re}\left( \lambda_5 \right)\  &
	\  \textrm{Im}\left( \lambda_5 \right)\  &
	\ - \textrm{Re}\left( \lambda_6 - \lambda_7 \right)\  \\*[1mm]
	\ \textrm{Im}\left( \lambda_6 + \lambda_7 \right)\  &
	\   \textrm{Im}\left( \lambda_5 \right)\  &
	\ - \lambda_4 + \textrm{Re}\left( \lambda_5 \right)\  &
	\ \textrm{Im}\left( \lambda_6 - \lambda_7 \right)\ \\*[1mm]
	\ -\tfrac{1}{2}(\lambda_1 - \lambda_2)\  &
	\ - \textrm{Re}\left( \lambda_6 - \lambda_7 \right)\  &
	\ \textrm{Im}\left( \lambda_6 - \lambda_7 \right)\  &
	\ - \tfrac{1}{2}(\lambda_1 + \lambda_2) + \lambda_3\
\end{array}
\right).
\label{Lambda}
\ee
Of course,
one can change the basis of the fields by
\be
\phi_i^\prime = U_{ij} \phi_j.
\ee
Henceforth,
using a roman lowercase letter in a four dimensional vector
refers implicitly to the spatial components:
$\mathbb{r}_k$, $k=1, 2, 3$.
Choosing the matrix $U$ in $U(2)$ guarantees that the kinetic
terms retain their canonical form.
In that case,
$\mathbb{r}_0$ is invariant and $\mathbb{r}_k$ suffers a $O(3)$ change.
However,
as {hinted in \cite{NagelPhD, Maniatis:2006fs}} and explored extensively in
Refs.~\cite{Ivanov:2006yq, Ivanov:2007de, Ivanov:2008cxa},
if one wishes to study exclusively the properties of the potential,
one can advantageously take any general transformation $U$ (unitary or not),
in which case  $\mathbb{r}^\mu$ suffers a $SO(1,3)$ change.
The advantage of these more general transformations
is that, for potentials bounded from below
(BFB) in the strong sense \cite{NagelPhD}, one can diagonalize $\Lambda_E$
into
\be
\Lambda_E
\rightarrow
\textrm{diag}
\left(\Lambda_0,\, \Lambda_1,\, \Lambda_2,\, \Lambda_3
\right),
\label{Lambda_diag}
\ee
and the eigenvalues satisfy \cite{Ivanov:2006yq}
\be
\Lambda_0 >0, \ \ \ \Lambda_0 > \Lambda_k.
\label{bfb}
\ee
Note that the signs of $\Lambda_k$ are arbitrary.

In the light of future discussions, it is instructive to see what changes within this formalism
if the potential is unbounded from below.
Although this situation is unphysical, one might accidentally run into it when
scanning over the parameter space.
In order to be sure that the discriminant we develop
below for the general 2HDM does not produce false positive results,
we must extend the bilinear formalism to this situation.

First, since $\Lambda_E$ is not symmetric, it might happen that some of its eigenvalues are complex.
There can be only one pair of complex and mutually conjugate eigenvalues,
see Appendix \ref{app:C}.
This leads to a potential unbounded from below,
as proved in the Erratum of Ref.~\cite{Ivanov:2006yq}.
Note that in this case we can still have a local minimum;
the simplest example being the inert 2HDM with $\lambda_2 < 0$.
Second, if all eigenvalues are real and some of them are degenerate, it is possible that
$\Lambda_E$ is not diagonalizable at all.
This can happen for a potential whose quartic part has flat directions,
so that it is not bounded from below in the strong sense.
Third, even if $\Lambda_E$ is diagonalizable \eqref{Lambda_diag},
it might happen that its eigenvalues do not satisfy \eqref{bfb}, in which case
the potential is also unbounded from below; yet a local minimum can exist.
Thus, a discriminating procedure must be guaranteed to cut off these cases
without referring to the diagonalization of $\Lambda_E$.

After spontaneous symmetry breaking (SSB),
the fields acquire vacuum expectation values:
\be
\langle \phi_1 \rangle
=
\frac{1}{\sqrt{2}}
\left(
\begin{array}{c}
	0\\
	v_1
\end{array}
\right),
\ \ \ \
\langle \phi_2 \rangle
=
\frac{1}{\sqrt{2}}
\left(
\begin{array}{c}
	0\\
	v_2\, e^{i \delta}
\end{array}
\right),
\label{vevs}
\ee
where, without loss of generality,
we have taken $\langle \phi_1 \rangle$ to be real.
Thus,
\be
r^\mu
=
\left( r_0, r_1, r_2, r_3 \right)
\equiv
\langle \mathbb{r}^\mu \rangle
=
\tfrac{1}{2}
\left(
v^2,\,
2 v_1 v_2 \cos{\delta},\,
2 v_1 v_2 \sin{\delta},\,
v_1^2 - v_2^2
\right),
\ee
where $v = \sqrt{v_1^2 + v_2^2} = (\sqrt{2} G_\mu)^{-1/2} = 246$ GeV.
Then,
$v_1$ and $v_2$ depend only on $\tan{\beta} = v_2/v_1$.
Notice that $\mathbb{r}^\mu$ refers to bilinears in fields,
while $r^\mu$ refers to their vevs.
Given our definitions,
the allowed $r^\mu$ satisfy
\be
r_0 \geq 0, \ \ \ r^\mu r_\mu \geq 0,
\ee
corresponding to the forward lightcone.
The $SU(2)\times U(1)$ symmetric vacuum lies at the apex,
the surface corresponds to neutral vacua,
while any point in the interior of the lightcone
represents charge breaking vacua.

\section{\label{sec:D}Discriminating the global minimum from metastable vacua}

Phenomenological analyses of 2HDM usually start with the
following procedure.
One assigns values to physical observables, such as masses,
mixing angles and couplings, which obey current (and simple) experimental
constraints.
For example,
one imposes that one of the Higgs masses equals 125 GeV.
The fact that masses squared are (of course) chosen
positive, implies that that point in parameter space
corresponds to a minimum; it may be local or global,
but it is guaranteed to be a minimum.
Further,
since either or both vevs are nonzero,
one is guaranteed to be in a minimum which breaks $SU(2) \times U(1)$.
But one is still unprotected against unpleasant situations where
one is sitting in the panic vacuum or when the
potential is unbounded from below.
From the input values, one extracts the parameters of the potential.
One can then impose theoretical constraints:
boundedness from below,
perturbative unitarity,
and avoidance of the panic vacuum.

When $\lambda_6 = \lambda_7 = 0$,
$\Lambda_E$ is block diagonal,
its eigenvalues can be found analytically,
and the conditions of bounded from below in the strong sense
can be written simply in terms of the parameters of the potential
as:
\be
\lambda_1 > 0,
\ \ \
\lambda_2 > 0,
\ \ \
\sqrt{\lambda_1 \lambda_2} > -\lambda_3,
\ \ \
\sqrt{\lambda_1 \lambda_2} > \left| \lambda_5 \right| - \lambda_3 - \lambda_4.
\label{bfb_L6L70}
\ee
When either $\lambda_6$ or $\lambda_7$ differ from zero \cite{Ivanov:2006yq},
one needs to construct the matrix $\Lambda_E$,
diagonalize it numerically
-- \textit{c.f.}\, Eqs.~\eqref{Lambda} and \eqref{Lambda_diag} --
and then impose Eqs.~\eqref{bfb}.
It turns out that,
for a general potential, this part of the procedure is more difficult than it seems.
The point is that, even after the diagonalization leading to
four eigenvalues $\Lambda_\alpha$ ($\alpha = 0, 1, 2, 3$),
one must identify which one of the four is $\Lambda_0$.
Ref~\cite{Barroso:2013awa} addressed this issue in the context of the CP conserving
two Higgs doublet model by resorting to the identification of
eigenvectors, a rotation matrix, and some rotated vectors.

Here we propose a procedure valid for the most general potential and
involving only the eigenvalues.
Imagine that we have identified the four eigenvalues $\Lambda_\alpha$,
that they are real,
and we wish to know which one to ascribe the subindex 0.
We construct the projection operators:
\be
\hat{P}^\alpha
=
\prod_{\beta \neq \alpha}
\frac{1}{\Lambda_\alpha - \Lambda_\beta}
\left( \Lambda_E - \Lambda_\beta  \mathbb{1} \right).
\label{projectors}
\ee
Then,
\be
s_\alpha =
\textrm{sign} \left[ (\hat{P}^\alpha )_{00} \right]
\label{s_alpha}
\ee
will be positive for only one value of $\alpha$;
the corresponding eigenvalue is the time-like $\Lambda_0$.
This assertion is proved in Appendix~\ref{app:A}.

The method proposed here to avoid panic vacua
uses exclusively the stationarity conditions and
the diagonalization of $\Lambda_E$
already needed to impose boundness from below.
The former can be obtained by minimizing the auxiliary potential
\be
\bar{V} = V - \tfrac{1}{2} \zeta\, r^\mu\, r_\mu,
\ee
with respect to $r^\mu$ and $\zeta$, yielding
\be
\Lambda^\mu_{.\, \nu}\, r^\nu - M^\mu = \zeta\, r^\mu.
\label{min}
\ee
Any component of this equation can be used to determine $\zeta$.
Typically,
equating $\zeta$ obtained from two different components
yields some $m_{ij}^2$ (in $M^\mu$) in terms of the quartic coefficients
$\lambda_k$ (in $\Lambda^\mu_{.\, \nu}$).
The quantity $\zeta$ calculated at any neutral stationary point
has a direct physical interpretation \cite{Maniatis:2006fs,Ivanov:2006yq}:
the charged Higgs mass squared is equal to $\zeta v^2$.
Our new discriminant is given by
\be
D = - \textrm{det} \left( \Lambda_E - \zeta \mathbb{1} \right),
\label{D}
\ee
where $\mathbb{1}$ is the four dimensional identity matrix.
Writing $\Lambda_E$ in the diagonal basis,
\be
D = (\Lambda_0 - \zeta)(\zeta - \Lambda_1)(\zeta - \Lambda_2)(\zeta - \Lambda_3).
\label{D_diag}
\ee
With these definitions, we introduce the following \textit{Method}:
\begin{enumerate}
	\item Determine $\zeta$ from Eq.~\eqref{min}.
	\item Determine $D$ from Eq.~\eqref{D}.
	\begin{enumerate}
		\item If $D >0$, then we are in the global minimum.
		\item If $D < 0$, then we must continue.
	\end{enumerate}
	\item Find the eigenvalues of $\Lambda_E$. If some of them are complex, discard the point. If all of them are real,
    use $s_\alpha$ in Eq.~\eqref{s_alpha} to identify $\Lambda_0$.
	\begin{enumerate}
		\item If $\Lambda_0 < \Lambda_k$ for some $k=1, 2, 3$, then the potential is not
		bounded from below and the point must be discarded.
		\item If $\Lambda_0 > \Lambda_k$ for all $k=1, 2, 3$, then we must continue.
		\begin{enumerate}
			\item If $\zeta > \Lambda_0$ then we are in a global minimum
            (there is no other minimum).
			\item If $\zeta < \Lambda_0$ then we are in the panic vacuum
			(metastable minimum).
		\end{enumerate}
	\end{enumerate}
\end{enumerate}
As mentioned,
in cases other than softly broken $Z_2$
the analysis of boundedness from below
requires the diagonalization of $\Lambda_E$,
while $D$ in Eq.~\eqref{D} can be calculated in any basis.
And, in many cases of interest it will turn out that our point has $D>0$.
This explains why we compute $D$ before checking whether the potential
is bounded from below.
When $D>0$, we are in the global minimum and we automatically know
that the potential is bounded from below.
There exists no other situation in which $D>0$ is compatible with a minimum.
This is proved in Appendices~\ref{app:B} and \ref{app:C}.
Thus, if $D>0$. metastability is avoided and the phenomenological analysis of
this point in parameter space can continue immediately.
The diagonalization of $\Lambda_E$ is only performed
when $D<0$.
This procedure solves completely the identification of the global minimum in the most
general 2HDM.

Since $\zeta = m_{H^\pm}^2/v^2$, we can also make another curious observation.
Suppose we know completely the quartic part of the potential and we know that it is bounded from below.
Then, the local vs. global minimum ambiguity is decided just by the value of the charged Higgs mass
through Eq.~\eqref{D}.

\section{\label{sec:examples}Application to a soflty broken $Z_2$ potential}

This section is dedicated to potentials with a
softly broken $Z_2$ symmetry,
where $\lambda_6=\lambda_7=0$ in Eq.~\eqref{VH}.
We allow $m_{12}^2$ to be complex (C2HDM)
or real (real 2HDM).
In addition, we may have $\lambda_5=0$ ($U(1)$-2HDM).
In such cases,
$\Lambda_E$ in Eq.~\eqref{Lambda}
is block diagonal and its eigenvalues are easy to find:
\ba
\Lambda_0 = \frac{1}{2} \left( \lambda_3 + \sqrt{\lambda_1 \lambda_2} \right),
&\ \ \ \ &
\Lambda_3 = \frac{1}{2} \left( \lambda_3 - \sqrt{\lambda_1 \lambda_2} \right),
\nonumber\\
\Lambda_1 = - \frac{1}{2} \left( \lambda_4 + |\lambda_5| \right),
&\ \ \ \ &
\Lambda_2 = - \frac{1}{2} \left( \lambda_4 - |\lambda_5| \right).
\label{Lambda_C2HDM}
\ea
This formulas lead from Eqs.~\eqref{bfb} to Eqs.~\eqref{bfb_L6L70}.
Moreover,
using a basis where both vevs are real\footnote{One can always find such a basis.
The phase will appear as part of $m_{12}^2$ and $\lambda_5$.},
we find
\ba
\zeta &=& - \frac{\lambda_4 + \textrm{Re}(\lambda_5)}{2}
+
\frac{\textrm{Re}(m_{12}^2)}{v_1 v_2}
\label{zeta_C2HDM}
\\
&=& - \frac{\lambda_4 + \lambda_5}{2}
+
\frac{m_{12}^2}{v_1 v_2}
\\
&=& \frac{m_{H^\pm}^2}{v^2}.
\ea
In going from the first to the second line we have used the
the stationarity conditions.
Some relevant additional formulas will be presented in
section~\ref{sec:C2HDM}.

Since both Eqs.~\eqref{bfb} and \eqref{D_diag} depend on
$\Lambda_\alpha$, there is an intimate connection
between boundedness from below and the discriminant for
panic vacua.
In cases other than soflty broken $Z_2$
(real or complex),
including in particular the most general CP conserving
potential,
there is no simplified compact form to assure that
the potential is bounded from below written in terms of
the original parameters of the potential $\lambda_1$--$\lambda_7$.
Thus,
as proposed in our \textit{Method} of section~\ref{sec:D},
it is best to start by computing $D$
(a determinant which can be calculated in any basis)
and only later perform the diagonalization of $\Lambda_E$
which enables the application of Eqs.~\eqref{bfb},
guaranteeing that the potential is bounded from below.
In contrast,
for soflty broken $Z_2$ potentials
($\lambda_6=\lambda_7=0$, $m_{12}^2$ real or complex),
the bounded from below conditions can be applied directly in the
form of Eqs.~\eqref{bfb_L6L70}.
In that case,
one can start by imposing these conditions and later study the discriminant
\ba
\tilde{D}
&=&
\frac{D}{\Lambda_0-\zeta}
=
 (\zeta - \Lambda_1)(\zeta - \Lambda_2)(\zeta - \Lambda_3)
\nonumber\\*[2mm]
&=&
\left[
\left(
\frac{m_{H^\pm}^2}{v^2} +
\frac{\lambda_4}{2}
\right)^2
- \frac{|\lambda_5|^2}{4}
\right]
\left[
\frac{m_{H^\pm}^2}{v^2}
+
\frac{\sqrt{\lambda_1 \lambda_2} - \lambda_3}{2}
\right]
.
\label{tildeD}
\ea
Then:

\begin{center}
\textit{We are in a global minimum if and only if $\tilde{D} > 0$.}
\end{center}

\noindent
We stress that the use of $\tilde{D}$ is only relevant \textit{after}
one has imposed boundedness from below.
This is simple for softly broken $Z_2$ models,
but requires the diagonalization of $\Lambda_E$ otherwise.
In those more general cases, one would be better served using
$D$ before diagonalization of $\Lambda_E$;
if $D>0$ the point corresponds to the global minimum
(the potential is guaranteed to be bounded from below)
and the phenomenological analysis may continue;
only if $D<0$ should the diagonalization of $\Lambda_E$ proceed
in order to determine whether the point corresponds to
a global minimum or not, along the lines of the
\textit{Method} of section~\ref{sec:D}

Using Eqs.~\eqref{Lambda_C2HDM} and \eqref{zeta_C2HDM},
it is easy to write both $D$ and $\tilde{D}$ in terms of the
parameters of the potential.
We will shortly use this to describe $D$ in terms of physical parameters.
However, the presence of the square root $\sqrt{\lambda_1 \lambda_2}$
in $\Lambda_0$ makes it impossible to write simple
expressions for $\tilde{D}$ in terms of physical parameters.
In contrast,
it is possible to write $D$ in terms of physical parameters.
This is what we turn to next.

\subsection{\label{sec:C2HDM}The C2HDM}

We consider here the C2HDM \cite{Ginzburg:2002wt, Khater:2003wq,
ElKaffas:2007rq, El Kaffas:2006nt, WahabElKaffas:2007xd,
Osland:2008aw, Grzadkowski:2009iz, Arhrib:2010ju, Barroso:2012wz},
for which $\lambda_6 = \lambda_7 =0$
and
$\textrm{arg}(\lambda_5) \neq 2 \textrm{arg}(m_{12}^2)$,
ensuring that there is explicit CP violation in the scalar potential.
We follow the notation of  Refs.~\cite{Fontes:2014xva, Fontes:2015mea}.
In particular, a basis is chosen such that both vevs are real --
$\delta=0$ in Eq.~\eqref{vevs}.
The minimization conditions are:
\ba
-2\, m_{11}^2 &=&
- 2\, \textrm{Re}\left(m_{12}^2 \right) \frac{v_2}{v_1}
+ \lambda_1\, v_1^2 + \lambda_{345}\, v_2^2
\nonumber\\
-2\, m_{22}^2 &=&
- 2\, \textrm{Re}\left(m_{12}^2 \right) \frac{v_1}{v_2}
+ \lambda_2\, v_2^2 + \lambda_{345}\, v_1^2
\nonumber\\
2\, \textrm{Im}\left(m_{12}^2 \right) &=& v_1 v_2\, \textrm{Im}\left(\lambda_5 \right),
\label{stat_cond}
\ea
where $\lambda_{345}= \lambda_3 +  \lambda_4 + \textrm{Re}\left(\lambda_5 \right)$.
We may parametrize the original fields as
\be
\phi_1 =
\left(
\begin{array}{c}
%\varphi_1^+
c_\beta G^+ - s_\beta H^+
\\
\tfrac{1}{\sqrt{2}} (v_1 + \eta_1 + i c_\beta G^0 - i s_\beta \eta_3)
\end{array}
\right),
\hspace{5ex}
\phi_2 =
\left(
\begin{array}{c}
%\varphi_2^+
s_\beta G^+ + c_\beta H^+
\\
\tfrac{1}{\sqrt{2}} (v_2 + \eta_2 + i s_\beta G^0 + i c_\beta \eta_3)
\end{array}
\right),
\ee
where $G^+$ and $G^0$ are the Goldstone bosons and $H^+$ is
charged Higgs with mass $m_{H^\pm}$.
Thus forth $c_\textrm{angle}$ ($s_\textrm{angle}$) refers to
the cosine (sine) of that angle.
The neutral mass matrix ${\cal M}^2$ is diagonalized by the orthogonal
transformation
\be
\left(
\begin{array}{c}
h_1\\
h_2\\
h_3
\end{array}
\right)
= R
\left(
\begin{array}{c}
\eta_1\\
\eta_2\\
\eta_3
\end{array}
\right),
\label{h_as_eta}
\ee
such that
\be
R\, {\cal M}^2\, R^T = \textrm{diag} \left(m_1^2, m_2^2, m_3^2 \right),
\ee
where the neutral scalar masses are ordered as $m_1 \leq m_2 \leq m_3$.
We parametrize the matrix $R$ by \cite{ElKaffas:2007rq}
\be
R =
\left(
\begin{array}{ccc}
c_1 c_2 & s_1 c_2 & s_2\\
-(c_1 s_2 s_3 + s_1 c_3) & c_1 c_3 - s_1 s_2 s_3  & c_2 s_3\\
- c_1 s_2 c_3 + s_1 s_3 & -(c_1 s_3 + s_1 s_2 c_3) & c_2 c_3
\end{array}
\right),
\ee
where $i$ is short for $\alpha_i$.
The limit of the real 2HDM is obtained by setting
$\alpha_2=\alpha_3=0$ and $\alpha = \alpha_1 - \pi/2$.
The parameters of the potential may be traded for physical observables through
\ba
v^2\, \lambda_1 &=&
- \frac{1}{\cos^2{\beta}}
\left[- m_1^2\, c_1^2 c_2^2 - m_2^2 (c_3 s_1 + c_1 s_2 s_3)^2
- m_3^2\, (c_1 c_3 s_2 - s_1 s_3)^2 + \mu^2\, \sin^2{\beta}
\right],
\nonumber\\*[2mm]
v^2\, \lambda_2 &=&
- \frac{1}{\sin^2{\beta}}
 \left[
- m_1^2\, s_1^2 c_2^2 - m_2^2\, (c_1 c_3 - s_1 s_2 s_3)^2
- m_3^2\, (c_3 s_1 s_2 + c_1 s_3)^2  + \mu^2\, \cos^2{\beta}
\right],
\nonumber\\*[2mm]
v^2\, \lambda_3 &=&
\frac{1}{\sin{\beta} \cos{\beta}}
\left[
\left(
m_1^2\, c_2^2
+ m_2^2\, (s_2^2 s_3^2 - c_3^2)
+ m_3^2\, (s_2^2 c_3^2 - s_3^2)
\right) c_1 s_1
\right.
\nonumber\\*[2mm]
& &
\hspace{10ex}
\left.
+\,
(m_3^2 - m_2^2) (c_1^2 - s_1^2) s_2 c_3 s_3
\right]
- \mu^2 + 2 m_{H^\pm}^2\, ,
\nonumber\\*[2mm]
v^2\, \lambda_4 &=&
m_1^2\, s_2^2 + ( m_2^2\,  s_3^2 + m_3^2\, c_3^2) c_2^2
+ \mu^2 - 2 m_{H^\pm}^2,
\nonumber\\*[2mm]
v^2\, \textrm{Re}(\lambda_5)
&=&
- m_1^2\, s_2^2 - (m_2^2\, s_3^2 + m_3^2\, c_3^2) c_2^2 + \mu^2,
\nonumber\\*[2mm]
v^2\, \textrm{Im}(\lambda_5)
&=&
\frac{2}{\sin{\beta}}
c_2
\left[
(- m_1^2 + m_2^2\, s_3^2 + m_3^2\, c_3^2) c_1 s_2
+ (m_2^2 - m_3^2) s_1 s_3 c_3
\right],
\label{reconstruct_lambdas}
\ea
with
\be
\mu^2 =
\frac{v^2}{v_1\, v_2}\, \textrm{Re}(m_{12}^2).
\label{eq:mu}
\ee
The observables are
$m_1$, $m_2$,
$m_{H^\pm}$, $\alpha_{1,2,3}$, $\beta$,
and $\textrm{Re}(m_{12}^2)$,
while $m_3$ is determined from
\be
m_3^2 = \frac{m_1^2\, R_{13} (R_{12} \tan{\beta} - R_{11})
+ m_2^2\ R_{23} (R_{22} \tan{\beta} - R_{21})}{R_{33} (R_{31} - R_{32} \tan{\beta})}.
\label{m3_derived}
\ee

We can now write the discriminant $D$ in terms of physical parameters.
Substituting Eqs.~\eqref{reconstruct_lambdas} into Eqs.~\eqref{Lambda_C2HDM}
and \eqref{zeta_C2HDM},
and those into Eq.~\eqref{D_diag},
we find
\be
D =
\frac{1}{8 v^8\, s_\beta^4\, c_\beta^2}\,
\left(- a_1\, \mu^2 + b_1\right)\, \left(a_2\, \mu^2 - 2\, b_2\right),
\ee
where
\ba
a_1 &=&
s_\beta^2
\left[
m_1^2 s_2^2 + \left( m_2^2 s_3^2 + m_3^2 c_3^2 \right) c_2^2
\right]\ ,
\nonumber\\
b_1 &=&
c_2^2
\left[
c_1 s_2 \left(
        -m_1^2 + m_2^2 s_3^2 + m_3^2 c_3^2
        \right)
+ s_1 s_3 c_3 \left( m_2^2 - m_3^2 \right)
\right]^2\ ,
\nonumber\\
a_2 &=&
2 m_1^2 c_2^2 c_{\alpha_1 + \beta}^2
+
\left( m_2^2 + m_3^2 \right)
\left( 1 - c_2^2 c_{\alpha_1 + \beta}^2 \right)
\nonumber\\
&&
+
\left( m_2^2 - m_3^2 \right)
\left[
\cos{(2 \alpha_3)}
\left( s_{\alpha_1 + \beta}^2 - c_{\alpha_1 + \beta}^2 s_2^2 \right)
+ \sin{(2 \alpha_3)} s_2 \sin\left( 2 \alpha_1 + 2 \beta \right)
\right]\ ,
\nonumber\\
b_2 &=&
\left( m_2^2 c_3^2 + m_3^2 s_3^2 \right) m_1^2 c_2^2
+ m_2^2 m_3^2 s_2^2 \ .
\ea

\subsection{\label{sec:real2HDM}The real 2HDM and the $U(1)$-2HDM}

We now consider a real softly broken $Z_2$ potential,
where $\lambda_6 = \lambda_7 = 0$,
and $\lambda_5$, $m_{12}^2$, and the vevs are real.
In this case,
CP is conserved,
$\eta_3=A$ is CP odd,
while the neutral CP even scalars are denoted
by $H$ (heavy) and $h$ (125 GeV).
In terms of the notation in the C2HDM,
the angles become $\alpha_2=\alpha_3=0$
and $\alpha= \alpha_1 - \pi/2$.
In this case,
the expression for $D$ in terms of physical parameters simplifies considerably
and can be programmed directly into any phenomenological analysis.
We find
\be
\frac{4 v^8 c_\beta^3 s_\beta^3}{m_A^2 m_h^2 m_H^2} D
=
m_{12}^2
\left[
1 - \frac{m_{12}^2}{m_h^2 m_H^2 s_{\beta}c_\beta}
(m_H^2 c^2_{\alpha + \beta} + m_h^2 s^2_{\alpha + \beta})
\right].
\ee

The discriminant introduced in reference
\cite{Barroso:2013awa} for the case
of the real 2HDM is
\be
D_{Z_2} =
m_{12}^2 (m_{11}^2 - k^2 m_{22}^2)(\tan{\beta} - k),
\label{DZ2}
\ee
where
\be
k = \sqrt[4]{\frac{\lambda_1}{\lambda_2}}.
\ee
This quantity can also not be
written directly in terms of physical parameters.
The problem,
as explained above, is related to the presence of
$\sqrt{\lambda_1 \lambda_2}$ in the $\Lambda_0$
of Eq.~\eqref{Lambda_C2HDM},
which is the origin of $k$ in Eq.~\eqref{DZ2}.

Although this expression is written in terms of the parameters of the potential and not the physical observables,
it can be recovered in our formalism in the following way.
We first calculate $D$ using (\ref{Lambda_C2HDM}),
\be
D = {m_{12}^2 \over v_1 v_2}{m_A^2 \over v^2} {k^4 (m_{22}^2)^2 - (m_{11}^2)^2 \over v_2^4 - k^4 v_1^4}\,,
\ee
and then relate it to $D_{Z_2}$ via
\ba
D &=&
\lambda_2^2 \frac{v_1^2}{v_2^2}
\frac{m_{12}^2 - \lambda_5 v_1 v_2}{(\lambda_1 v_1^4 - \lambda_2 v_2^4)^2}
(- m_{11}^2 - k^2 m_{22}^2)
(\tan{\beta} + k)(\tan^2{\beta} + k^2)
D_{Z_2}\,,
\label{D_and_DZ2}
\ea
where each factor is positive definite.
Combining the pre-factors in Eqs.~\eqref{DZ2} and \eqref{D_and_DZ2},
we that $D$ depends only on $k^4$ and, thus, we understand why it
can be written in terms of physical parameters in a simpler way.

Things get even simpler in the case of the potential with a
softly broken $U(1)$ symmetry.
The discriminant for this case was introduced in Ref.~\cite{Barroso:2012mj}
as
\be
D_{U(1)} =
(m_{11}^2 - k^2 m_{22}^2)(\tan{\beta} - k).
\label{DU1}
\ee
For this model,
$2 m_{12}^2 = m_A^2 s_{2 \beta} >0 $, and the discriminant may be taken as
\begin{equation}
\frac{4 v^8 c_\beta^2 s_\beta^2}{m_A^4 m_h^2 m_H^2} D
=
1 - \frac{m_A^2}{m_h^2 m_H^2}
(m_H^2 c^2_{\alpha + \beta} + m_h^2 s^2_{\alpha + \beta}).
\end{equation}

\section{\label{sec:concl}Conclusions}

When studying models with two Higgs doublets,
it is possible that the vacuum chosen is not the global one,
and that there is another vacuum lying below.
In that situation, the vacuum is metastable,
there could be a later transition into the global vacuum,
and we dub this situation the \textit{panic vacuum}.
We have developed a method to avoid panic vacua,
involving the discriminant $D$ in Eqs.~\eqref{D}-\eqref{D_diag},
which is applicable to any 2HDM potential.
In particular, we proved that $D>0$ guarantees that we are staying in the global minimum
and, automatically, the potential is bounded from below.
If $D < 0$, more steps are needed to discriminate between
the global minimum from a panic vacuum or a potential unbounded from below.
Our method is not computer time consuming
and it be easily implemented in phenomenological
studies in which extensive scans over 2HDM parameter space.

We have shown how $D$ can be written in terms of physical parameters
for the C2HDM, the real 2HDM and the $U(1)$-2HDM,
and we have shown how it is related with the discriminants
previously presented for the later two cases
\cite{Barroso:2012mj, Barroso:2013awa}.

If the potential has a softly broken $Z_2$ symmetry,
a simpler strategy is possible.
Indeed, in that case,
the bounded from below conditions are easy to implement
-- \textit{c.f.} Eq.~\eqref{bfb_L6L70}.
Then one need only apply $\tilde{D}$ in Eq.~\eqref{tildeD}
and \textit{one has a global minimum if and only if $\tilde{D} > 0$}.

This article completes the identification of panic vacuua
for all 2HDM, including the previous unsolved cases where
CP violation is present in the scalar potential.

\begin{acknowledgments}
J.P.S. is grateful to J. Rom\~{a}o for discussions.
This work is supported in part by the Portuguese
\textit{Funda\c{c}\~{a}o para a Ci\^{e}ncia e Tecnologia} under contract
UID/FIS/00777/2013.
I.P.I. acknowledges funding from the \textit{Funda\c{c}\~{a}o para a Ci\^{e}ncia e Tecnologia}
through the FCT Investigator contract IF/00989/2014/CP1214/CT0004
under the IF2014 Programme.
\end{acknowledgments}

\appendix

\section{\label{app:A}Proof that $s_0 > 0$}

The operators $\hat{P}^\alpha$ introduced in Eq.~\eqref{projectors},
one for each eigenvalue $\Lambda_\alpha$ of $\Lambda_E$,
obey
\be
\hat{P}^\alpha \hat{P}^\beta = \delta_{\alpha \beta} \hat{P}^\alpha,
\ \ \ \ \ \
\sum_{\alpha} \hat{P}^\alpha = \mathbb{1},
\ee
and  $\hat{P}^\alpha$ is the projection operator into the subspace
generated by the eigenvector corresponding to $\Lambda_\alpha$.
Since our method uses projectors after we have discarded complex eigenvalue situation,
we know that all $\Lambda_\alpha$ are real.

The explanation of the proof is easiest in Minkovski
notation, where
\be
(\hat{P}^\alpha)_{\mu \nu}
=
\prod_{\beta \neq \alpha}
\frac{1}{\Lambda_\alpha - \Lambda_\beta}
\left( \Lambda_{\mu \nu} - \Lambda_\beta g_{\mu \nu} \right),
\ee
which are symmetric in the indices $\mu$, $\nu$.
Take now some four-vector $\ell^\mu$ (for example $\ell^\mu=(1,0,0,0)$).
The new vector
\be
\ell_\nu^0 = (\hat{P}^0)_{\nu \mu} \ell^\mu
\ee
lies along the eigenvector of $\Lambda_0$,
and, thus, it is timelike:
\be
0 < (\ell^0)^\nu \ell_\nu^0 =
\ell^{\mu^\prime}
(\hat{P}^0)_{\mu^\prime}^{\, .\, \nu}
(\hat{P}^0)_{\nu \mu} \ell^\mu
=
\ell^{\mu^\prime}
(\hat{P}^0)_{\mu^\prime \mu}
\ell^\mu.
\ee
Similarly,
\be
\ell_\nu^k = (\hat{P}^k)_{\nu \mu} \ell^\mu
\ee
for $k=1,2,3$ lies along the eigenvector of $\Lambda_k$,
and, thus, it is spacelike:
\be
0 > (\ell^k)^\nu \ell_\nu^k
=
\ell^{\mu^\prime}
(\hat{P}^k)_{\mu^\prime \mu}
\ell^\mu.
\ee
This is true for any vector $\ell^\mu$, apart, of course, from the
case when the vector is accidentally chosen to be orthogonal to some eigenvector.
Choosing the simplest case $\ell^\mu=(1,0,0,0)$,
means that
\be
s_\alpha = \textrm{sign} \left[ (\hat{P}^\alpha )_{00} \right]
\ee
is positive if and only if $\alpha=0$.
This completes our proof.

\section{\label{app:B}Locating stationary points in parameter space}

Imagine that $\Lambda_0$, $\Lambda_k$, and $\zeta$ have been identified.
Some ordering of $\Lambda_\alpha$ has been found.
If $\Lambda_0$ is the largest,
we denote the spacelike eigenvalues by
$\Lambda_a < \Lambda_b < \Lambda_c$.
In this appendix we show how one can determine how many
extrema exist, depending on where $\zeta$ sits with respect
to the $\Lambda_\alpha$.
We follow the analysis of Refs.~\cite{NagelPhD, Maniatis:2006fs, Ivanov:2007de},
also used in \cite{Barroso:2013awa} in the particular case $\hat{M}_2=0$.

The stationarity conditions \eqref{min} may be written in the
basis where $\Lambda_E$ is diagonal as
\be
(\Lambda_0 - \zeta) \hat{r}_0 = \hat{M}_0,
\ \ \ \
(\Lambda_k - \zeta) \hat{r}_k = \hat{M}_k,
\label{min2}
\ee
where the hat in $\hat{r}$ and $\hat{M}$ emphasize the fact that
these vectors are written in the basis where $\Lambda_E$ is diagonal.
Let us introduce the variables
\be
n_k = \frac{\hat{r}_k}{\hat{r}_0}\,,\quad
m_k = \frac{\hat{M}_k}{\hat{M}_0}\,,\quad
a_k(\zeta) = \frac{| \Lambda_k - \zeta|}{|\Lambda_0 - \zeta|}\,,
\ee
where $n_k$ is a variable unit vector, $m_k$ is a fixed vector,
and $a_k(\zeta)$ are just numbers which depend on $\zeta$.
The system \eqref{min2} then takes the following form:
\be
a_k(\zeta)\, n_k = m_k\,.
\label{min3}
\ee
No summation is assumed here.
Take some fixed $\zeta$ and all possible directions of the unit
vector $n_k$.
Then, the left hand side of the system of Eqs.~\eqref{min3}
defines (the surface of) an ellipsoid with semiaxes $|a_k(\zeta)|$.
As $\zeta$ changes, this ellipsoid grows and shrinks in a way that can be well visualized.
Whenever this ellipsoid crosses the fixed point $(m_1, m_2, m_3)$,
the system \eqref{min3} is satisfied, and we get an extremum.
Depending on the relations among $\Lambda_\alpha$ and $\zeta$,
as well as on the location of $m_k$, this can happen at most six times,
yielding up to six stationary points.

\begin{figure}[h!]
	\centering
	\includegraphics[width=0.3\linewidth]{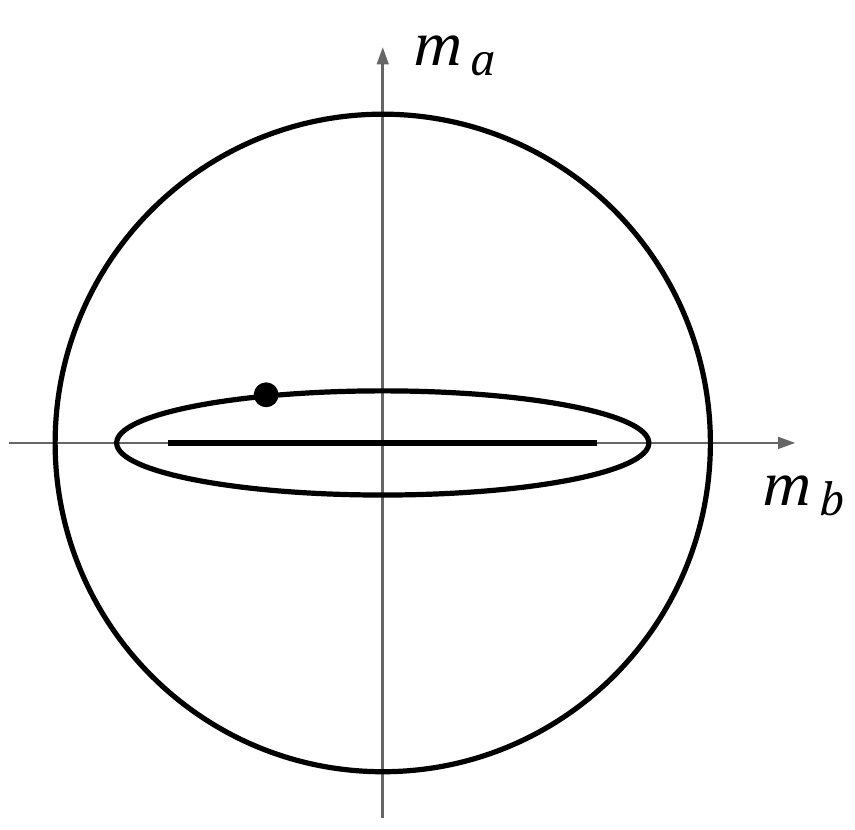}
	\hspace{0.02\linewidth}
	\includegraphics[width=0.3\linewidth]{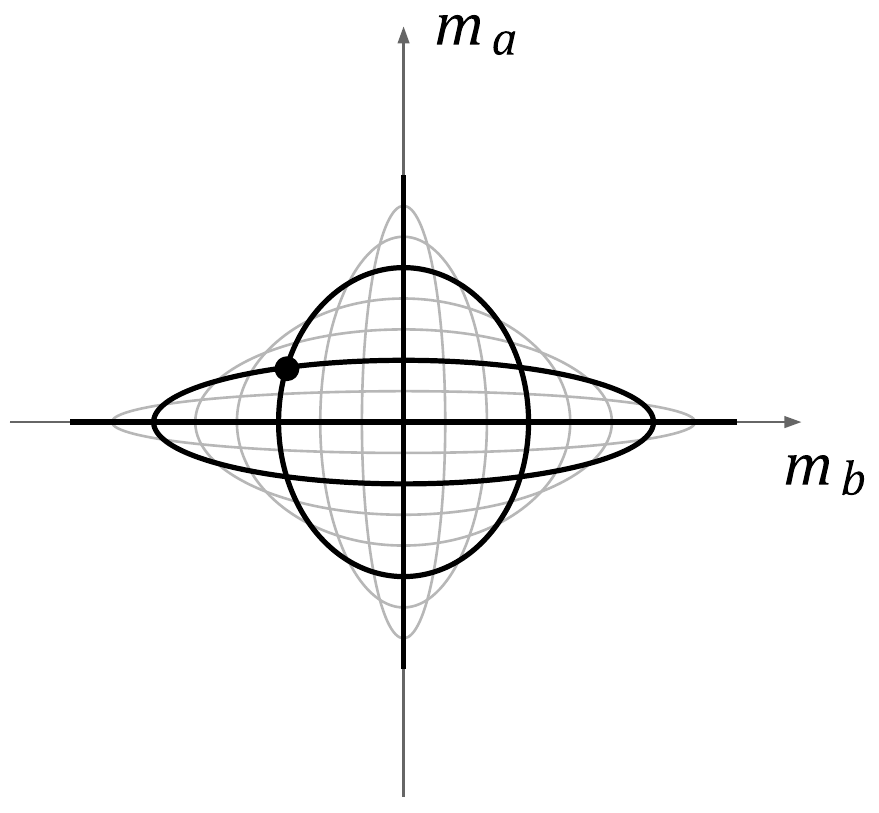}
	\caption{Projection on the $(m_a, m_b)$ plane of the ellipsoid given
		by Eq.~\eqref{ellipsoid} as $\zeta$ varies from $-\infty$ to
		$\Lambda_a$ (left panel) and from $\Lambda_a$ to $\Lambda_b$
		(right panel).
		The black dot represents a fixed point
		on the $(m_a, m_b)$ plane}
	\label{fig:1}
\end{figure}

Although these properties are reasonable to assume given the
work of Refs.~\cite{NagelPhD, Maniatis:2006fs, Ivanov:2007de},
we motivate them here for completeness.
We start by noting that
$\sum_k n_k^2 = 1$ and Eqs.~\eqref{min3} lead to
\be
\frac{m_a^2}{a_a(\zeta)^2} +
\frac{m_b^2}{a_b(\zeta)^2} +
\frac{m_c^2}{a_c(\zeta)^2}
= 1\, ,
\label{ellipsoid}
\ee
which, for some fixed $\zeta$ can be viewed as the surface of
an ellipsoid in the $(m_a, m_b, m_c)$ space.
We will now see what happens to this surface
as $\zeta$ increases from $-\infty$ to $+ \infty$.
Recall that we have ordered the eigenvalues such that
$\Lambda_a < \Lambda_b < \Lambda_c$.
In the limit that $\zeta \rightarrow - \infty$,
we obtain the unit sphere.
As $\zeta$ increases, the sphere becomes an ellipsoid whose
semi-axis along $m_a$ shrinks faster than the rest.
When $\zeta = \Lambda_a$, $a_a(\zeta)=0$ and the ellipsoid becomes
a filled ellipse on the $(m_b, m_c)$ plane, flat along $m_a$.
This process is illustrated on the left panel of Fig.~\ref{fig:1},
through a projection on the $(m_a, m_b)$ plane.
%
%%%%%%%%%%%%%%%%%%%%%%%%%%%%%%%%%%%%%%%%%%%%%%%%%%%%%%%%%%%%%%%%%%%%%%%%%
%%%%%%%%%%%%%%%%%%%%%%%%%%%%%%%%%%%%%%%%%%%%%%%%%%%%%%%%%%%%%%%%%%%%%%%%%%%
%
Consider a fixed point on the $(m_a, m_b, m_c)$ space.
For simplicity,
we illustrate this with a point on the ${m_a, m_b}$ plane,
shown as a black dot in Fig.~\ref{fig:1}.
During the process depicted in on the left panel of Fig.~\ref{fig:1}
that point is crossed only once,
and, thus,
Eq.~\eqref{min3} is satisfied only once.
This means that there can be at most one stationary point
for a $\zeta$ in this region.

Now consider that $\zeta \in [\Lambda_a, \Lambda_b]$.
As $\zeta$ moves away from $\Lambda_a$,
$a_a(\zeta)$ increases again from zero,
while $a_b(\zeta)$ decreases.
This process is illustrated on the right panel of Fig.~\ref{fig:1},
through a projection on the $(m_a, m_b)$ plane.
When $\zeta = \Lambda_b$ we reach $a_b(\zeta)=0$,
and the ellipsoid collapses onto the $(m_a, m_c)$ plane.
This is shown as the vertical line
on the right panel of Fig.~\ref{fig:1}.
Along this process, some points may be passed twice
(as happens, for example, for the illustrated by the black dot),
meaning that one can have at most two stationary points.

The analysis for $\zeta \in [\Lambda_b, \Lambda_c]$ follows the
same lines and, again,
one concludes that there can be at most  two stationary points
in this region.
Finally,
taking $\zeta \in [\Lambda_c, \Lambda_0]$,
the ellipsoid grows without bound and all points in the $(m_a, m_b, m_c)$ space
are crossed exactly once.
As shown in
Refs.~\cite{NagelPhD, Maniatis:2006fs},
larger $\zeta$ correspond to smaller
values for the potential.
In our notation, if $r^\mu$ and $r^{\prime\mu}$ are
two stationary points with the corresponding values $\zeta > \zeta'$,
then
\ba
|V(r)| - |V(r')| = {1\over 2}(\zeta-\zeta')r^\mu r'_\mu > 0\,.
\ea
This means that the largest $\zeta$ for which Eq.~\eqref{min3}
holds will be the global minimum
(provided the potential is bounded from below).
In the case discussed here,
this would occur for $\zeta \in [\Lambda_c, \Lambda_0]$.

Due to the positivity of $\hat r_0$, the above description holds for $\hat M_0 > 0$.
If $\hat M_0 < 0$, then $\zeta$ can vary from $\Lambda_0$ to $+\infty$.
If $\zeta$ starts from $+\infty$ and decreases, then the ellipsoid starts from the unit sphere
and monotonously grows to infinity as $\zeta$ approaches $\Lambda_0$.
This process covers exactly once all points in the $(m_a, m_b, m_c)$ space
outside the unit sphere.
We get a single extremum in this case, which must be the global minimum.

Finally, if it happens that $\Lambda_0$ is not the largest among the eigenvalues,
the above construction is still valid up to obvious modifications
and can be used to count extrema in each region.
However the potential in this case is not bounded from below,
and this picture does not allow one to spot the presence of this fact.

\section{\label{app:C}Proof that $D > 0$ guarantees a global minimum}

When building a 2HDM potential, we know by construction
that we are at a minimum, and we just want to know whether this minimum
is global. In our method, we first provide a discriminant $D$, whose positive sign
guarantees that we are at a global minimum. Once again we repeat that
$D>0$ is a sufficient but not necessary condition for the global minimum,
and our method contains extra steps to be checked in the case of $D<0$.

Here, we prove that this is indeed a sufficient condition.
We do it by showing that all other possible stationary points with $D>0$
do not correspond to a minimum. Since our method suggests checking $D$
even before we diagonalize $\Lambda_E$, this claim must cover all possible cases,
including potentials unbounded from below.\bigskip

{\bf Non-pathological case.}
Let us first build the proof for the case when all eigenvalues of $\Lambda_E$ are real
and this matrix is diagonalizable.
To distinguish a minimum from a saddle point,
one must consider the Hessian
\ba
H_{ab} \equiv {1\over 2}{\partial^2 V \over \partial
\varphi_a \partial \varphi_b}\, ,
\label{hessian}
\ea
where for clarity the Higgs doublets are rewritten in terms of eight real fields $\varphi_a$, $a=1,\dots,8$.
One then checks that, apart from would-be Goldstone modes,
the Hessian is positive definite in the Higgs field space.
For a neutral stationary point, the charged and neutral fields decouple,
and one focuses
on $H_{ab}$ in the four-dimensional subspace of neutral Higgs modes
(labeled by $a,b=1\dots4$).

\begin{figure}[h!]
	\centering
	\includegraphics[width=0.75\linewidth]{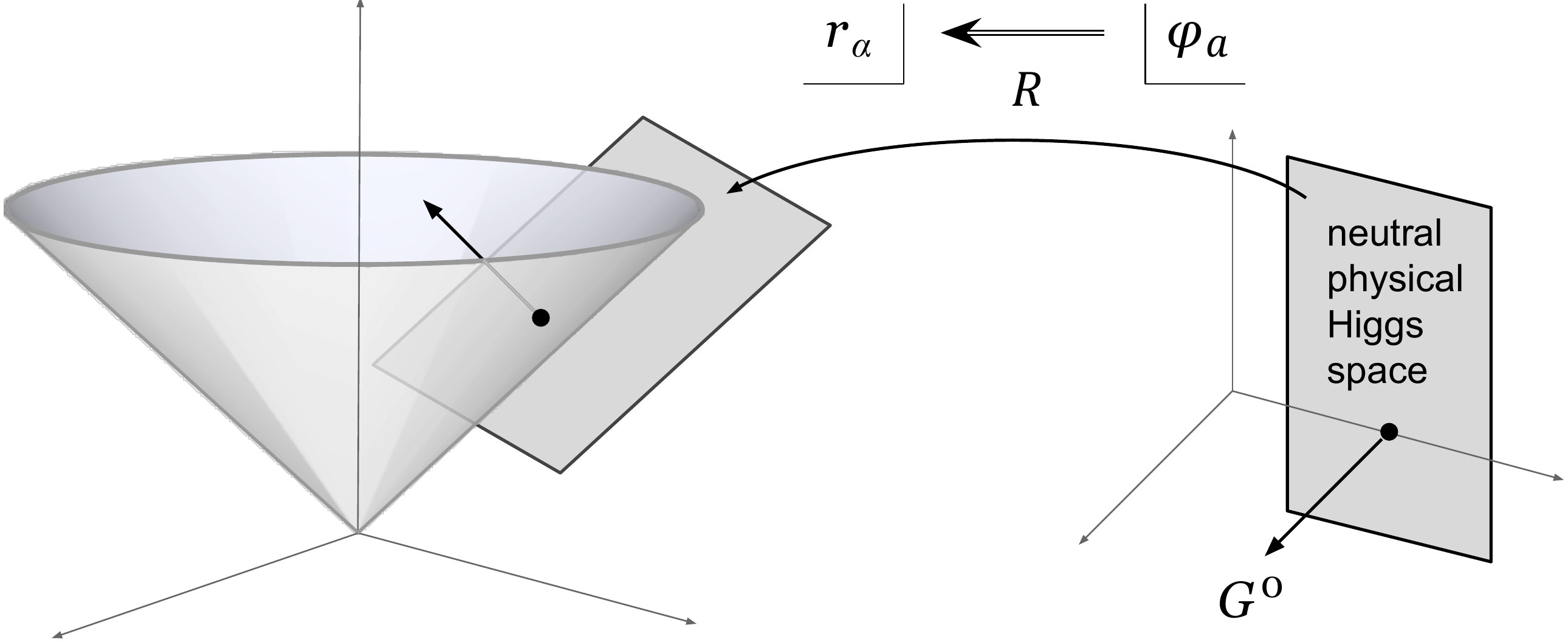}
	\caption{The correspondence between the 4D spaces established by the
		transformation $R$ in \eqref{hessian-2}.
		The neutral field space contains a would-be Goldstone $G^0$,
		which is annihilated by $R$.
		Its complement, the physical Higgs space, shown in gray,
		is mapped onto a 3D plane tangent to the forward
		lightcone in the $r_\alpha$ space at the extremum point.
		The direction orthogonal to this plane points inside the
		cone and it cannot be represented by any neutral scalar mode.}
	\label{fig:2}
\end{figure}

Within the bilinear approach, the neutral Higgs mass matrix
was written in the Higgs basis in \cite{Maniatis:2006fs,Ivanov:2006yq,Nishi:2007dv}.
Despite being compact,
those expressions do not provide insight on how the value of $\zeta$
with respect to $\Lambda_0$ and $\Lambda_k$ is related with minimum
versus saddle point assignments.
To gain it, we use instead the basis-invariant approach
to Higgs masses \cite{Degee:2009vp}, which allows us to switch
to the $\Lambda_{\mu\nu}$-diagonal basis.
In this basis, the Hessian
in the 4D space of neutral modes takes the following form:
\be
H_{ab} = (R^T)_{a\alpha} S_{\alpha \beta} R_{\beta b}\,,
\ee
where
\ba
S
&=&
\left(
\begin{array}{cccc}
\Lambda_0 - \zeta & 0 & 0 & 0 \\
0 & \zeta - \Lambda_1 & 0 & 0 \\
0 & 0 & \zeta - \Lambda_2 & 0  \\
0 & 0 & 0 & \zeta - \Lambda_3
\end{array}
\right)\,,
\nonumber\\*[2mm]
R_{a \alpha}
&=&
{1\over 2}
\left\langle
{\partial \mathbb{r}_\alpha \over \partial \varphi_a}
\right\rangle\,.
\label{hessian-2}
\ea
There are a few important remarks concerning this formula.
First, all indices here refer to four dimensional Euclidean spaces,
and $H_{ab}$ can be calculated as a usual product of three matrices.
However, there are two different spaces involved here.
The indices $\alpha, \beta$ refer to the same
space of bilinears from which we erased its Minkowski space metric,
while the indices $a, b$ refer to the space of neutral scalar modes at the extremum.
These two 4D spaces are shown in Fig.\ref{fig:2}.
%
%
%%%%%%%%%%%%%%%%%%%%%%%%%%%%%%%%%%%%%%%%%%%%%%%%%%%%%%%%%%%%%%%%%%%%%%%%%
%%%%%%%%%%%%%%%%%%%%%%%%%%%%%%%%%%%%%%%%%%%%%%%%%%%%%%%%%%%%%%%%%%%%%%%%%%%
%
%
Second,
the signature matrix $S$ and the matrix $\Lambda_E - \zeta 1$,
whose determinant we calculate to get $D$,
originate from the same $\Lambda_{\mu\nu} - \zeta g_{\mu\nu}$:
the former is obtained by erasing the Minkowski metric;
the latter is obtained by lowering one index.
This leads to $D = \det S$.
However,
as we will see below,
in our discussion we will need not only $\det S$ but also its full signature;
defined as the number of positive and negative eigenvalues.
Third,
since the diagonalization of $\Lambda_{\mu\nu}$ belongs
in general, to $SO(1,3)$,
it modifies the kinetic term of the Higgs fields,
and the mass matrix is not equal to the Hessian \eqref{hessian-2}.
However such a transformation,
effectively rotating and stretching the Higgs space,
leaves invariant the signature of the Hessian
(the signs of its eigenvalues) -- see proposition 1 in
Ref.~\cite{Ivanov:2006yq}.
Therefore,
the signature of $H_{ab}$ faithfully represents the signs of
the masses squared of the four neutral scalar degrees of freedom.

The link between the signatures of $H_{ab}$ and of $S$ is less trivial.
The transformation matrix, $R$, is singular: $\det R = 0$,
which indicates that one of the four directions
in the scalar field space $\varphi_a$ is, in fact,
a would-be Goldstone boson.
Orthogonal to it lies the 3D physical neutral Higgs space
(thenceforth referred to as the ``physical Higgs space''),
which is mapped by $R$ onto the 3D subspace in the $r_\alpha$
space tangent to the lightcone
at the extremum point.
Therefore, when distinguishing a minimum from a saddle point,
one must pay attention not to the full signature of $S$
but to the signature of its restriction
onto this 3D subspace.

This allows us to establish the following relation between the
signatures of $S$ and $H_{ab}$.
\begin{itemize}
\item
If $S$ has the signature $(+,+,+,+)$,
then $H_{ab}$ is positive definite in the physical Higgs space,
and we are at a minimum. Also, $D>0$.
\item
If $S$ has the signature $(+,+,+,-)$,
up to permutations,
then $H_{ab}$ is not positive definite in the entire
$r_\alpha$ space,
but it might still be positive definite when restricted to
the physical Higgs space.
In this case $D<0$,
but deciding whether we are at a minimum or at a saddle point
requires further analysis.
\item
If $S$ has the signature $(+,+,-,-)$,
up to permutations,
then $H_{ab}$ cannot be positive definite even when projected
onto the physical Higgs space.
This results from dimension counting:
in the $r_\alpha$ space, there exists a 2D subspace of negative $S_{\alpha\beta}r_{\alpha}r_{\beta}$,
which must intersect a 3D subspace tangent to the lightcone.
So, this stationary point cannot be a minimum but it has $D>0$.
The same conclusion holds for the signature $(-,-,-,-,)$.
\end{itemize}
With this classification in mind,
a minimum yielding positive $D$ can take place if
and only if $S$ has signature $(+,+,+,+)$.
This, in turn, can happen only for $\Lambda_0 > \zeta > \Lambda_c$,
the largest among $\Lambda_k$,
which automatically implies that the potential is bounded from below.
In the previous subsection,
we established that there can exist only one stationary point
in this region of $\zeta$,
and that this stationary point must be a minimum.
In this situation, there cannot be any other stationary point
with $\zeta > \Lambda_0$.
Therefore, a minimum with $D>0$ is the global minimum.\bigskip

{\bf Pathological case.}
Now we turn to the case of one pair of complex
and mutually conjugate eigenvalues of $\Lambda_E$.
Now, $\Lambda_{\mu \nu}$ cannot be diagonalized with any transformation
from the group $SO(1,3)$.
Nevertheless, it can be brought, with the aid of an $SO(1,3)$ transformation, to the block-diagonal form
\be
\Lambda_E =
\left(
\begin{array}{cccc}
\Lambda_{00} & -\Lambda_{01} & 0 & 0 \\
\Lambda_{01} & - \Lambda_{11} & 0 & 0 \\
0 & 0 & \Lambda_2 & 0  \\
0 & 0 & 0 & \Lambda_3 \\
\end{array}
\right)\,,
\ee
where $\Lambda_2$ and $\Lambda_3$ are the two real
eigenvalues of $\Lambda_E$.
It can still be used to calculate $D = - \det(\Lambda_E - \zeta 1)$.
The expression for the Hessian (\ref{hessian-2})
is also valid with the following signature matrix $S$:
\be
S =
\left(
\begin{array}{cccc}
\Lambda_{00} -\zeta & \Lambda_{01} & 0 & 0 \\
\Lambda_{01} & \Lambda_{11} +\zeta & 0 & 0 \\
0 & 0 & \zeta -\Lambda_2 & 0  \\
0 & 0 & 0 & \zeta- \Lambda_3 \\
\end{array}
\right)\,.
\ee
Now, we have the following sequence of arguments.
Consider the discriminant $D$ on the entire $\zeta$ axis.
By construction, it changes sign only
when $\zeta$ passes through an eigenvalue,
which in this case happens only at $\zeta=\Lambda_2$
and $\zeta=\Lambda_3$.
Since $D < 0$ at $\zeta \to \infty$,
we find that the only region where $D$ is positive
is between $\Lambda_2$ and $\Lambda_3$.
The upper block in $\Lambda_E$ gives the following strictly positive factor to
the discriminant
$(\zeta-\Lambda_{00})(\zeta+\Lambda_{11}) + \Lambda_{01}^2 > 0$
(otherwise, the eigenvalues on this subspace would be real,
contrary to our assumption).
The corresponding upper block in $S$ has an extra minus sign
in its second row and, therefore,
it contributes one ``$-$'' entry to the signature of $S$.
Since $\zeta$ is between $\Lambda_2$ and $\Lambda_3$,
we get a second ``$-$'' from this subspace.
Overall, the signature matrix $S$ possesses two negative eigenvalues
and cannot correspond to a minimum.

\end{document}